\documentclass[
   ,final            
  ,draft            
  ,numberedheadings 
  ]{aipproc}

\layoutstyle{8x11single}

\usepackage{amsmath, amsthm, amssymb}

\newcommand{\dif}{\mathrm{d}}

\begin{document}
\vspace*{20mm}
\tiny
\title 
      [Fluctuations of CMBR]
      {Fluctuations of CMBR in accelerating universe}
\classification{98.70.Vc}
\keywords      {Cosmology, fluctuations of CMBR, accelerating universe, Rees-Sciama effect}

\author{Zden\v{e}k  Stuchl\'{i}k}{
  address={Institute of Physics, Faculty of Philosophy and Science, Silesian University at Opava\\Bezru\v{c}ovo n\'{a}m. 13, 74601 Opava, Czech Republic},
  email={zdenek.stuchlik@fpf.slu.cz},
  thanks={This work is supported by the Czech grand MSM 4791305903}
}

\author{Jan Schee}{
  email={schee@email.cz},
}

\copyrightyear  {2005}

\begin{abstract}
\footnotesize
 The influence of the observed relict vacuum energy on the fluctuations of CMBR going 
through cosmological matter condensations is studied in the framework
 of the Einstein-Strauss-de Sitter vakuola model. It is shown that refraction of light
 at the matching surface of the vakuola and the expanding Friedman universe can be very important
 during accelerated expansion of the universe, when the velocity of the matching surface relative
 to static Schwarzchildian observers becomes relativistic. Relevance of the refraction
 effect for the temperature fluctuations of CMBR is given in terms of the redshift and
 the angular extension of the fluctuating region.
 \end{abstract}

\maketitle


\section*{Introduction}\label{intro}
\small
Temperature fluctuations of the Cosmic Microwave Background Radiation (CMBR),
recently measured by COBE, WMAP, etc., are observed on the level of $\Delta T/T \sim 10^{-5}$
\cite{Spe-etal:2003::astro-ph/0302209}. These fluctuations can be explained
in two ways. First, by the Sachs--Wolfe effect \cite{Sac-Wol:1967:ASTRJ2:},
i.e., as an imprint of energy density fluctuations related to the CMBR
temperature fluctuations at the cosmological redshift $z\sim 1300$ during the
era of recombination, when effective interaction of matter and CMBR is ceased
\cite{Bor:1993:EarlyUniv:}.  Second, by the Rees--Sciama effect
\cite{Ree-Sci:1968:NATURE:}, i.e., as a result of influence of large-scale
inhomogeneities (large galaxies or their clusters, and large voids)
evolved in the expanding universe due to the gravitational instability of
matter at the era characterised by $z\lesssim 10$. In the case of spherically
symmetric clusters and voids, the Rees--Sciama effect was considered in detail
by M\'esz\'aros and Moln\'ar~\cite{Mes-Mol:1996:ASTRJ2:}. They describe the
clusters by the standard Einstein--Strauss vakuola model, while the voids they
model in an approximative way that does not meet the full general-relativistic
junction conditions. Further, they do not consider the effect of refraction of
light at the boundary surface matching the cluster (void) with the expanding
universe.  However, this effect could be of great importance in an
accelerating universe, indicated by many of recent cosmological tests
predicting present value of the vacuum energy density $\rho_{\mathrm{vac}}\sim
0.7\rho_{\mathrm{crit}}$ ($\rho_{\mathrm{crit}}\equiv 3H/8\pi G$ is the
critical energy density corresponding to the flat universe predicted by the
inflationary
paradigm~\cite{Lin:1990:InfCos:,Spe-etal:2003::astro-ph/0302209}). The vacuum
energy density (or energy of a quintessence field) is related to the
(effective) cosmological constant by
\begin{equation}
  \Lambda = \frac{8\pi G}{c^2} \rho_{\mathrm{vac}}.         \label{rce_I1} 
\end{equation}

\begin{figure}[t]
\includegraphics[width=.45\linewidth]{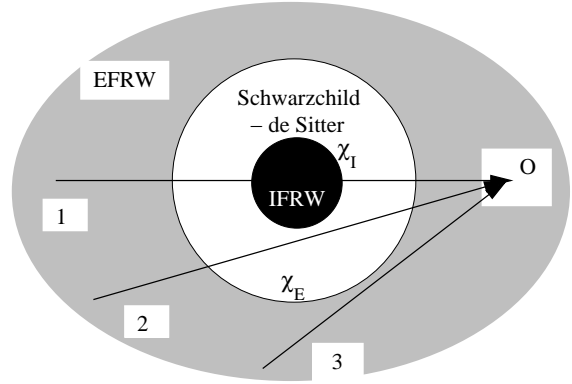}
\caption{ESdS vakuola model. A schematic picture of a cluster represented as a
sphericaly symmetric inhomogeneity immersed in the dust filled Friedman universe. 
The observer O receives two photons coming through the vakuola, and third one coming
 directly.\label{obr1}}
\end{figure}

Here, we present a study of the influence of the relict repulsive cosmological
constant, indicated by observations to be equal $\Lambda \approx
10^{-56}\,\mathrm{cm^{-2}}$, on the Rees--Sciama effect. We use the
Einstein--Strauss--de~Sitter (ESdS) vakuola model in which the inhomogeneity is
represented by a spherically symmetric cluster immersed into the
Friedmanian dust-filled universe (see Fig.\,\ref{obr1}). We determine
temperature fluctuations of the CMBR passing the vakuola described by the
ESdS model and give estimations of the relevance of
the effect of refraction at the matching surface.

We use the geometric units with $c=G=1$.

\section{Einstein--Strauss--de~Sitter vakuola model}\label{ESdSvm}

In the construction of the ESdS model with a $\Lambda>0$, we remove a spherical ball of dust of
the mass $M$ from the dust-filled universe and replace it by the
Schwarzschild--de~Sitter spacetime of the same mass $M$. Its expanding boundary
surface coincides at a fixed value of the comoving Robertson-Walker (RW) coordinate $\chi_E=\chi_{\mathrm{b}}$ 
with expanding surface $\chi = \chi_{\mathrm{b}} = \mathrm{const}$ of the Friedman universe (see Fig.\ref{obr1}). 
The Schwarzschild--de~Sitter (SdS) spacetime can be completely vacuum, i.e., a black-hole spacetime, or, as used
frequently, it has a spherical source represented by a part of an internal
dusty Friedman universe with parameters different than those of the external
Friedman universe outside of the vacuum SdS spacetime, and characterized by $\chi_I<\chi_E$. For simplicity, we shall not 
consider influence of the source of the internal part of the ESdS vakuola model on the CMBR fluctuations.

In the standard Schwarzschild coordinates, the vacuum SdS spacetime of mass $M$ is described by the line element
\begin{equation}
  \dif s^2 = - \mathcal{A}^2(r)\,\dif t^2
             + \mathcal{A}^{-2}(r)\,\dif r^2
             + r^2\,\dif\Omega,\quad   \mathcal{A}^2(r) = 1-\frac{2M}{r}-\frac{\Lambda}{3}r^2.      \label{rce2}
\end{equation}

The Friedman universe is described by the FRW geometry. In the comoving coordinates its line element reads
\begin{equation}
  \dif s^2 = - \dif T^2
             + R^2(T)\left[\dif\chi^2
                     +\Sigma^2_k(\chi)\,\dif\Omega\right],     \label{rce3}
\end{equation}
where
\begin{equation}
  \Sigma_k(\chi)=
    \left\{
           \begin{array}{rll}
              \sin\chi & \mbox{for} & k=+1,\\
                  \chi & \mbox{for} & k=0,\\
             \sinh\chi & \mbox{for} & k=-1.
           \end{array}     
    \right.                                                    \label{rce4}
\end{equation}
The RW metric describes the external Friedman universe at
$\chi\ge\chi_E=\chi_{\mathrm{b}}$, while at $\chi < \chi_{\mathrm{b}}$ it is replaced
by the expanding part of the SdS spacetime. The particles
with $\chi=\chi_{\mathrm{b}}$ follow radial geodesics of the
SdS spacetime.

The evolution of the Friedman universe is given by the evolution of the scale
factor $R$ and the energy density $\rho$ in dependence on the cosmic time $T$.
The scale factor fulfils the Friedman equation
\begin{equation}
  \left(\frac{\dif R}{\dif T}\right)^{2} = \frac{8\pi\rho}{3R}
    + \frac{\Lambda}{3}R^2 - k                                 \label{rce5}
\end{equation}
and the energy density $\rho$ satisfies the energy conservation equation in
the form
\begin{equation}
  \frac{8\pi\rho}{3}R^3 = \mathrm{const} = R_0.                \label{rce6}
\end{equation}

It is necessary to synchronize the proper time of a dust particle on the
matching hypersurface (MH hereinafter) $\chi=\chi_{\mathrm{b}}$ as measured
from the both sides of the MH. Therefore, the proper
time $\tau$ of a test particle following the radial geodesic, as measured in the
SdS spacetime, must be equal to the cosmic time $T$, as
measured in the FRW spacetime. The junction conditions
have the following form~\cite{Stu:1983:BULAI:}
\begin{equation}
  r_{\mathrm{b}} = R(T)\,\Sigma_k(\chi_{\mathrm{b}}),\quad \tilde R = R_0\Sigma_k(\chi_{\mathrm{b}}),\quad \tilde R \sqrt{\tilde R/2M} = R_0,\label{rce8}
\end{equation}
where the parameter $\tilde R$ is related to the covariant energy $\mathcal{E}_{\mathrm{b}}$ of the test particles along the
radial geodesic, representing the MH by the relation
\begin{equation}
  \mathcal{E}_{\mathrm{b}} = \sqrt{1-\frac{2kM}{\tilde R}}.   \label{rce10}
\end{equation}
The internal 3-geometry of the MH as measured from the FRW universe side is
given by the line element
\begin{equation}
  \dif s_+^2 = - \dif T^2
               + R^2(T)\,\Sigma^2_k(\chi_{\mathrm{b}})
    \left(\dif\theta^2 + \sin^2\theta\,\dif\phi^2\right).     \label{rce11}
\end{equation}
From the side of the SdS spacetime it is given by the line
element
\begin{equation}
  \dif s_-^2 = - \dif T^2
               + r_{\mathrm{b}}^2(T)
  \left(\dif\theta^2 + \sin^2\theta\,\dif\phi^2\right).       \label{rce12}
\end{equation}
Both geometries are identical due to the junction conditions. One can show
that the same statement holds for the extrinsic curvature of the
MH~\cite{Stu:1984:BULAI:}.

\section{The geodesics intersecting the matching hypersurface}\label{igeo}%

Let us consider geodesics crossing the MH\@. We have to find the relation
between the directional angle as measured by the comoving Friedman observers,
$\psi_{\mathrm{F}}$, and the directional angle as measured by the
Schwarzchild--de Sitter static observers,$\psi_{\mathrm{S}}$. The segments of the geodesics in the FRW and the SdS
geometry must be smoothly connected on the MH\@. We are looking for the
Lorentz transformation relating the comoving Friedman and the static
SdS observers on the MH\@.

In the RW metric, the geodesic equations can be integrated and
expressed in the form~\cite{Stu:1984:BULAI:}
\begin{equation}
  p^T=\frac{\dif T}{\dif \lambda}
    = \left(m^2 + \frac{p^2}{R^{2}}\right)^{1/2},\quad 
  p^\chi=\frac{\dif\chi}{\dif \lambda}
    =\pm\frac{1}{R^{2}}
     \left(p^2-\frac{L^2}{\Sigma_k^2}\right)^{1/2},          \label{rce_R2}
\end{equation}

\begin{equation}
  p^\theta=\frac{\dif\theta}{\dif\lambda}
    =\pm\frac{1}{R^{2}\Sigma_k^2}
     \left(L^2+\frac{\ell^2}{\sin^2\theta}\right)^{1/2},\quad 
  p^\phi=\frac{\dif\phi}{\dif \lambda}
    =\frac{\ell}{R^{2}\Sigma_k^2\sin^2\theta},               \label{rce_R4}
\end{equation}
where $\lambda$ is an affine parameter and $m$ is mass of the particle; the
proper time $\tau = m\lambda$. The constants of motion are
\begin{equation}
  \ell=p_\phi,\quad                                             
  L^2=p^2_\theta + \frac{p^2_\phi}{\sin^2\theta},\quad          
  p^2=p^{2}_\chi + \frac{L^2}{\Sigma_k^2},                 \label{rce_R7}
\end{equation}
where $\ell(L)$ represent the azimuthal (total) angular momentum. Geodesic
equations in the SdS spacetime are in the integrated form
expressed by the formulae
\begin{equation}
  p^t=\frac{\dif t}{\dif\lambda} = E\mathcal{A}^{-2} (t),\quad 
  p^r=\frac{\dif r}{\dif \lambda}
    =\pm\left(E^2 -V^{2}_{\mathrm{eff}}\right)^{1/2},       \label{rce_R88}
\end{equation}
\begin{equation}
  p^\theta=\frac{\dif\theta}{\dif\lambda}
    =\pm\frac{1}{r^2}
      \left(L^2+\frac{\ell^2}{\sin^2\theta}\right)^{1/2},\quad  
  p^\phi=\frac{\dif\phi}{\dif\lambda}
    =\frac{\ell}{r^2\sin^2\theta},                           \label{rce_R9}
\end{equation}
where the effective potential
\begin{equation}
  V^2_{\mathrm{eff}}
    =\mathcal{A}^2(r)\left(m^2 + \frac{L^2}{r^2}\right).     \label{rce_R10}
\end{equation}
The constants of motion $\ell$ and $L$ have the same meaning as in the FRW case. The covariant energy $E=-p_t$.
\par
Let us consider coordinate systems with coincidentally oriented coordinate
axes, moving mutually in the direction of the radial axis. The orthonormal base
vectors are related by the standard Lorentz transformation
\begin{equation}
  \mathbf{e}_{(\mu^\prime)}
    = \Lambda_{\mu^\prime}^{\hphantom{\mu^\prime}\nu}
      \mathbf{e}_{(\nu)}                                    \label{rce_R12}
\end{equation}
with the Lorentz matrix
\begin{equation}
  \Lambda^{\hphantom{\mu^\prime}\nu}_{\mu^\prime}
    = \left(
      \begin{array}{cccc}
        \cosh{\alpha} & \sinh{\alpha} & 0 & 0\\
        \sinh\alpha & \cosh\alpha & 0 & 0\\
        0 & 0 & 1 & 0\\
        0 & 0 & 0 & 1
      \end{array}
      \right).                                              \label{rce_R13}
\end{equation}

The orthonormal basis of the static SdS observers is given
by the relations
\begin{equation}
  \mathbf{e}_{(t)}
    =\mathcal{A}^{-1}(r)\frac{\partial}{\partial t},\quad      
  \mathbf{e}_{(r)}
    =\mathcal{A}(r)\frac{\partial}{\partial r},           \label{rce_R15}
\end{equation}
\begin{equation}
  \mathbf{e}_{(\theta)}
    =r^{-1}\frac{\partial}{\partial\theta},\quad              
  \mathbf{e}_{(\phi)}
    = (r\sin\theta)^{-1}\frac{\partial}{\partial\phi},     \label{rce_R17}
\end{equation}
while in the case of the comoving FRW observers it is given by
\begin{equation}
  \mathbf{e}_{(T)}
    = \frac{\partial}{\partial T},\quad                        
  \mathbf{e}_{(\chi)}
    = R^{-1}\frac{\partial}{\partial\chi},                \label{rce_R19}
\end{equation}
\begin{equation}
  \mathbf{e}_{(\theta)}
    = (R\Sigma_k)^{-1}\frac{\partial}{\partial\theta},\quad    
  \mathbf{e}_{(\phi)}
    = (R\Sigma_k\sin\theta)^{-1}
      \frac{\partial}{\partial\phi}.                        \label{rce_R21}
\end{equation}
We obtain the parameter of the Lorentz transformation from the fact that the
4-velocity of the test particles comoving with the MH can be expressed in the
FRW and SdS spacetimes by the relation
\begin{equation}
  \mathbf{u}_{\mathrm{(b)}}
    = \frac{\partial}{\partial T}
    =\mathbf{e}_T                                                 
    =\mathcal{A}^{-1}(r_{\mathrm{b}})\mathcal{E}_{\mathrm{b}}\mathbf{e}_t
      + \left[\mathcal{E}^2_{\mathrm{b}} -
              \mathcal{A}^2(r_{\mathrm{b}})
        \right]^{1/2}
      \mathcal{A}^{-1}(r_{\mathrm{b}})\mathbf{e}_r.         \label{rce_R22}
\end{equation}
Therefore, we arrive at the Lorentz transformation parameter in the form
\begin{equation}
  \cosh\alpha
    =\Lambda_T^{\hphantom{T}t}=\Lambda_\chi^{\hphantom{\chi}r}
    =\mathcal{E}_{\mathrm{b}}\mathcal{A}^{-1}(r_{\mathrm{b}})     
    =\sqrt{1-\frac{2kM}{\tilde R}}
    \left(
      1-\frac{2M}{r_{\mathrm{b}}}
      -\frac{\Lambda r^2_{\mathrm{b}}}{3}
    \right)^{-1/2}.                                         \label{rce_R23}
\end{equation}
The velocity parameter of the Lorentz shift, giving the speed of the expansion of the MH as measured by the static
SdS observers, and the Lorentz factor are then given by the relations

\begin{equation}
  V(r_{\mathrm{b}})
    = \sqrt{1-\frac{\mathcal{A}^2(r_{\mathrm{b}})}%
      {\mathcal{E}^2_{\mathrm{b}}}},\quad                            
  \gamma = \cosh\alpha
    = \left[1-V(r_{\mathrm{b}})^2\right]^{-1/2}.            \label{rce_R25}
\end{equation}

\section{Refraction of light at the matching hypersurface}\label{refli}

Denoting the directional angles (related to the outward radial direction
defined for observers at the radius, where the MH is located momentarily) of a
photon entering (leaving) the FRW universe from (into) the
SdS vakuola as $\psi_{\mathrm{F}}^+$,
$\psi_{\mathrm{S}}^+$ ($\psi_{\mathrm{F}}^-$, $\psi_{\mathrm{S}}^-$), we
arrive at the formulae
\begin{equation}
  \cos\psi_{\mathrm{F}}^\pm
    =\frac{\cos\psi_{\mathrm{S}}^\pm \mp V_r}%
            {1\mp V_r\cos\psi_{\mathrm{S}}^\pm},\quad\quad                     
  \sin\psi_{\mathrm{F}}^\pm = \frac{\sin\psi_{\mathrm{S}}^\pm\sqrt{1 - V_r^2}}{1\mp V_r\cos\psi_{\mathrm{S}}^\pm}.\label{R_1}
\end{equation}

\begin{figure}[htb]
    \includegraphics[width=1.0\textwidth]{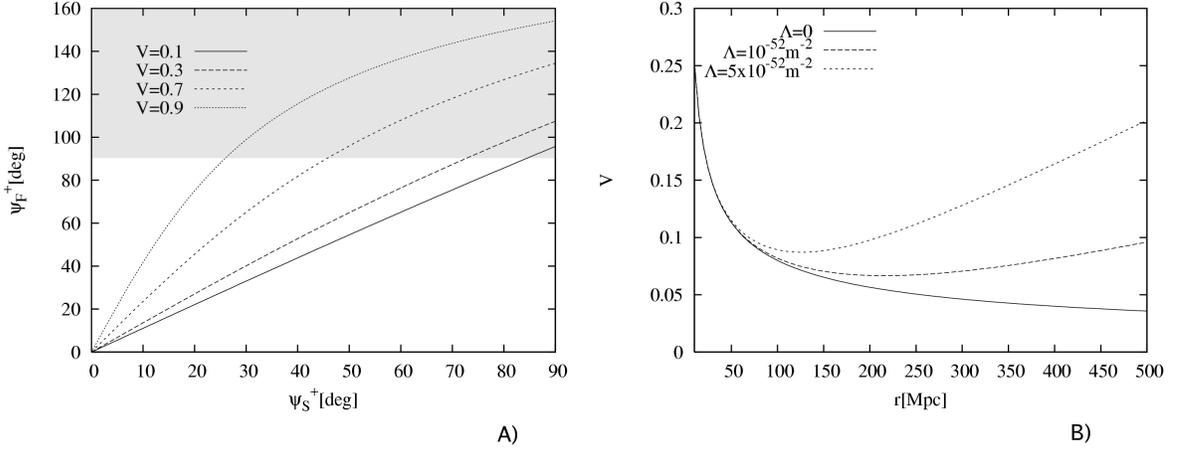}
    \caption{Plot A: The refraction angle $\psi_F^+$ for fixed speed parameter $V$, as a function
of the impact angle $\psi_S^+$ from the interval $[0, \pi/2]$. The shaded area
corresponds to the total reflection of the light. Plot B: The dependence $V_r=V_r(r_{\mathrm{b}})$ for fixed mass $M=7\cdot 10^{18}M_{sol}$ of vakuola and three representative
values of $\Lambda=0$, $10^{-52} m^{-2}$ and $5\cdot 10^{-52} m^{-2}$ and vakuola radius $r_{\mathrm{b}}$ from the interval
$[10Mpc,500Mpc]$. The function $V(r_\mathrm{b})$ reaches its local minimum as it approaches the static radius $r_{\mathrm{s}}$. \label{obr2}}
\end{figure}

For photons entering the FRW universe from the SdS spacetime, the detailed analysis \cite{Stu-Sch:2004:} shows that

\begin{equation}
  \psi_{\mathrm{F}}^+ > \psi_{\mathrm{S}}^+ \quad\mbox{for}\quad
    \psi_{\mathrm{S}}^+\in[0,\pi/2],                           \label{R_21}
\end{equation}
i.e., for such photons the refraction angle is always larger than the impact angle. The total reflection occurs for angles
\begin{equation}
  \psi_{\mathrm{S}}^+>\psi_{\mathrm{S(T)}}^+\equiv\arccos{V_r}.
\end{equation}
In Table~\ref{tab1}, we give the critical angles of the total refraction
$\psi_{\mathrm{S(T)}}^+$ for some values of the MH expansion velocity. In Fig.\,\ref{obr2}a, we present the dependence
$\psi_{\mathrm{F}}^+=\psi_{\mathrm{F}}^+(\psi_{\mathrm{S}}^+;V_r)$ for some
appropriately chosen values of the expansion velocity.

\begin{table}[t]
\caption{Total reflection angle $\psi^+_{\mathrm{S(T)}}$, calculated for four
  different values of the speed parameter $V_r$.\label{tab1}}
\begin{tabular}{lcccc}
\hline
$V_r$ & $0.1$ & $0.3$ & $0.7$ & $0.9$ \\
\hline 
$\psi^+_{\mathrm{S(T)}}$
  & $84^\circ 15^\prime$ & $72^\circ 32^\prime$ & $45^\circ 34^\prime$
  & $25^\circ 50^\prime$\\
\hline
\end{tabular}
\end{table}

Considering photons entering the SdS region from FRW universe, we can conclude that $\psi_{\mathrm{S}}^->\psi_{\mathrm{F}}^-$ for
$\psi_{\mathrm{F}}^-\in[0,\pi/2]$, i.e., the refraction angle
is again always larger then the impact angle, and the total reflection occurs
for
\begin{equation}
  \psi_{\mathrm{F}}^->\psi_{\mathrm{F(T)}}^-\equiv\arccos{V_r}.
\end{equation}

\subsection{The expansion velocity of the matching hypersurface}

We shall consider the simplest case of the MH expansion velocity for the spatially flat universe ($k=0$):
\begin{equation}
  V_r = \sqrt{\frac{2M}{r_{\mathrm{b}}}
    +\frac{\Lambda r_{\mathrm{b}}^2}{3}}.                      \label{R_25}
\end{equation}
Introducing a dimensionless cosmological parameter $y\equiv\frac{1}{3}\Lambda M^2$, we find the local extremum of $V_r(r_{\mathrm{b}})$
($\dif{V_r}/\dif{r_{\mathrm{b}}}=0$) located at so called static radius of the SdS spacetime
\begin{equation}
  \frac{r_{\mathrm{s}}}{M}\equiv y^{-1/3},                               \label{R_30}
\end{equation} 
where the gravitational attraction of the central mass condensation (or a black
hole) is just balanced by the cosmic
repulsion~\cite{Stu-Hle:1999:PHYSR4:}. We can see that with $r_b$ growing $V_r(r_{\mathrm{b}})$
falls down for $r_{\mathrm{b}}<r_{\mathrm{s}}$, it reaches its minimum at the
static radius ($r_{\mathrm{b}}=r_{\mathrm{s}}$), where
\begin{equation}
  V_{\mathrm{r(min)}}=V_r(r_{\mathrm{b}}
    =r_{\mathrm{s}})=\frac{3M}{r_{\mathrm{s}}}=3y^{1/3},        \label{R_31}
\end{equation}
while the expansion speed is accelerated at $r_{\mathrm{b}}>r_{\mathrm{s}}$,
approaching velocity of light ($V_r\rightarrow 1$) when $r_{\mathrm{b}}$
approaches the cosmological horizon of the SdS
region ($r_{\mathrm{b}}\rightarrow r_{\mathrm{c}}$) (see Fig.\ref{obr2}b). Notice that for $y\ll 1$,
the cosmological horizon is approximately given by
\begin{equation}
  \frac{r_{\mathrm{c}}}{M} \sim y^{1/2}.                                 \label{R_32}
\end{equation}
For the exact formulae giving loci of the event horizons $r_{\mathrm{c}}$ and $r_{\mathrm{h}}$ in
the SdS spacetimes see~\cite{Stu-Hle:1999:PHYSR4:}.

\section{Influence of the refraction effect on temperature fluctuations of the
  CMBR}

\begin{figure}[t]
\includegraphics[width=.5\linewidth]{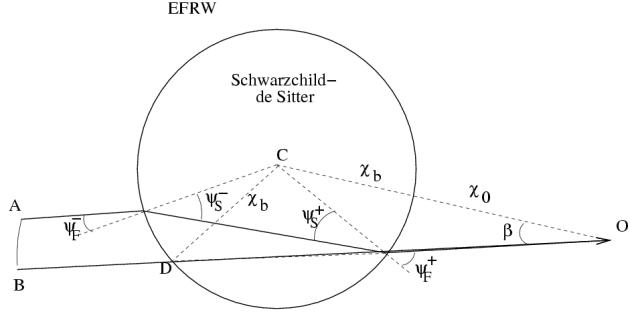}
\caption{Refraction of a photon ray going through the vakuola.
  $\chi_{\mathrm{b}}$ is the comoving coordinate of the vakuola boundary,
  $\chi_0$ is the comoving `radial' distance of the observer from vakuola,
  $\chi = \chi_{\mathrm{b}} + \chi_0$. The ray $BO$, with no refraction effect
  considered in accord with \protect\cite{Mes-Mol:1996:ASTRJ2:}, is included
  for comparison with previous results in order to clear up the relevance of
  the refraction on the Rees--Sciama effect. \label{obr3}}
\end{figure}

We shall study the influence of the refraction effect on the CMBR in the
framework of the ESdS model using the simplified
approach developed by M\'esz\'aros and Moln\'ar (for more detailed model,
considering also deflection of light by the mass condensation,
see~\cite{Dye:1973:ASTRJ2:}). We do not consider the model of void used
in~\cite{Mes-Mol:1996:ASTRJ2:}, since it is not self-consistent from the point
of view of general relativity. It was shown in~\cite{Mes-Mol:1996:ASTRJ2:}
that the temperature fluctuations are fully determined by the length of the
photon ray spanned in the vakuola region, i.e., it is determined by the angle
$\psi_{\mathrm{S}}$ giving the impact angle of photon on the MH. The effect of 
refraction can be incorporated into
the model by substituting the angle
$\psi_{\mathrm{S}}^+$ influenced by the refraction effect directly into the
formula determining the temperature fluctuation. For simplicity, we shall
consider here photon trajectories which do not enter the internal Friedman
region, and, as usual in the model, we abandon deflection of light in the
SdS spacetime. The impact angle $\psi_{\mathrm{S}}^+$ then
has to be related to the view angle $\beta$ of observer through the angle of
refraction $\psi_{\mathrm{F}}^+$ (see Fig.\,\ref{obr3}).

The temperature fluctuation (frequency shift) of a CMBR photon due to
transversing the vakuola is given by the
relation~\cite{Mes-Mol:1996:ASTRJ2:}
\begin{equation}
  \Delta T = \frac{2c^3Y^3}{H^3}
    \left\{\frac{\Omega}{2}\sin^2\psi\cos\psi +
    \frac{1+2\Omega}{3}\cos^3\psi\right\},                     \label{TF_1}
\end{equation}
where $\psi=\psi_{\mathrm{S}}^+$ determines the length of the ray in the
vakuola; $Y=R(\eta)\chi$ is the actual physical extension of the vakuola
and $H=\dot R/R$ is the actual value of the Hubble parameter; $R(\eta)$
is the scale factor, $\dot R\equiv\dif R/\dif T$, $\eta$ is the conformal time
defined by $d\eta=dT/R$.

Refraction effect will change the length of light ray spanning the vakuola
region (see Fig.\,\ref{obr3}). Of course, for vanishing refraction effect, there is $\psi_{\mathrm{S}}^+=\psi_{\mathrm{F}}^+$
in agreement with \cite{Mes-Mol:1996:ASTRJ2:}. Using formula (\ref{TF_1}) we find the temperature fluctuation with the refraction 
effect included to be given by the relation
\begin{eqnarray}
  \Delta T_r &=& \frac{2c^{3}Y^{3}}{H^{3}}  
    \left[
      \frac{\cos\psi_{F}^{+}+V(r_{\mathrm{b}})}%
           {1+V(r_{\mathrm{b}})\cos\psi_{F}^{+}}
    \right] \nonumber\\  
  &\times&
    \left\{\frac{\Omega}{2}
      \left[
        \frac{\sin\psi_{F}^{+}}%
             {\gamma\left(1+V(r_{\mathrm{b}})\cos\psi_{F}^{+}\right)}
      \right]^{2}
     + \frac{1+2\Omega}{3}
      \left[
        \frac{\cos\psi_{F}^{+}+V(r_{\mathrm{b}})}%
             {1+V(r_{\mathrm{b}})\cos\psi_{F}^{+}}
      \right]^{2}
    \right\}.                                                  \label{TF_5}
\end{eqnarray}

The relevance of the refraction effect is given by the difference of the
temperature fluctuations $\Delta T_r$ and $\Delta T$:
\begin{eqnarray}
  \Delta\equiv\Delta T_r - \Delta T &=& \frac{2c^3Y^3}{H^3}\left\{\frac{\Omega}{2}
      \left[
        \frac{\cos\psi_{\mathrm{F}}^+ + V_r}%
             {1+V_r\cos\psi_{\mathrm{F}}^+}
          \left(
            \frac{\sin\psi_{\mathrm{F}}^+}%
                 {\gamma[1+V_r\cos\psi_{\mathrm{F}}^+]}
          \right)^2
          -\cos\psi_{\mathrm{F}}^+\sin^2\psi_{\mathrm{F}}^+
      \right]
    \right.                                                       \nonumber\\
    &+&
    \left.\frac{1+2\Omega}{3}
      \left[
        \left(
          \frac{\cos\psi_{\mathrm{F}}^+ + V_r}%
               {1+V_r\cos\psi_{\mathrm{F}}^+}
        \right)^3
        -\cos^3\psi_{\mathrm{F}}^+
      \right]
    \right\}.                                                   \label{TF_6}
\end{eqnarray}
\noindent 
In the limit of non-relativistic velocities, $V_r\ll 1$, the relations
(\ref{R_1}) imply
\begin{equation}
  \cos\psi_{\mathrm{S}}^+
   \sim\cos\psi_{\mathrm{F}}^+ + V_r\sin^2\psi_{\mathrm{F}}^+,  \qquad
  \sin\psi_{\mathrm{S}}^+\sim
    \sin\psi_{\mathrm{F}}^+(1-V_r\cos\psi_{\mathrm{F}}^+),     \label{TF_8}
\end{equation}
so that up to the first order of $V_r$, the temperature difference is given by
the formula
\begin{equation}
  \Delta\equiv\Delta T_r - \Delta T\sim
    \frac{2c^3Y^3}{H^3}
    V_r\cos^2\psi_{\mathrm{F}}^+\sin^2\psi_{\mathrm{F}}^+
    \left(
      1+\Omega+\frac{\Omega}{2}\tan^2\psi_{\mathrm{F}}^+
    \right).                                                   \label{TF_9}
\end{equation}
Clearly, as we expected intuitively, the influence of the refraction effect
vanishes linearly with $V_r\rightarrow 0$.

The relevance of the refraction effect in terms of the
viewing angle $\beta$ follows directly from the sine rule (see Fig.\,\ref{obr3})
\begin{equation}
  \sin\psi_{\mathrm{F}}^+
    =\frac{\chi_0+\chi_{\mathrm{b}}}%
          {\chi_{\mathrm{b}}}\sin\beta                       \label{TF_10}
\end{equation}
\noindent
and from the relation between the Schwarzschild coordinate $r_{\mathrm{b}}$, and
the Robertson--Walker comoving coordinate $\chi_{\mathrm{b}}$ given by
\begin{equation}
  r_{\mathrm{b}}=R(t_{\mathrm{b}})\chi_{\mathrm{b}}=
    \frac{R_0}{1+z}\chi_{\mathrm{b}},                         \label{TF_11}
\end{equation}
where $R_0$ is recent value of $R$, and $z$ is the cosmological redshift,
being the measure of the cosmic time.  Introducing new variables
\begin{equation}
  A(\beta)
    =\frac{
      \sqrt{1-
        \left(
          \frac{\chi_0+\chi_{\mathrm{b}}}{\chi_{\mathrm{b}}}\sin\beta
        \right)^2}
      +V_r}%
      {1+V_r
        \sqrt{1-
          \left(
            \frac{\chi_0+\chi_{\mathrm{b}}}%
                 {\chi_{\mathrm{b}}}\sin\beta
          \right)^2}},\quad                                        
  B(\beta)
    =\frac{\frac{\chi_0 +\chi_{\mathrm{b}}}{\chi_{\mathrm{b}}}\sin\beta}%
            {\gamma
              \left[
                1+V_r
                  \sqrt{1-\left(\frac{\chi_0+\chi_{\mathrm{b}}}%
                                     {\chi_{\mathrm{b}}}\sin\beta
                          \right)^2}
              \right]},                                       \label{TF_13}
\end{equation}	    

\begin{equation} 
  C(\beta)
    =
      \sqrt{1-
        \left(
          \frac{\chi_0+\chi_{\mathrm{b}}}{\chi_{\mathrm{b}}}\sin\beta
        \right)^2}
        \left(
          \frac{\chi_0+\chi_{\mathrm{b}}}{\chi_{\mathrm{b}}}\sin\beta
        \right),                                              \label{TF_14}
\end{equation}
the temperature difference (\ref{TF_6}) can be expressed as a function of
$\beta$ in the form
\begin{equation}
  \Delta T_r - \Delta T
    = \displaystyle\frac{2c^3Y^3}{H^3}
      \left\{
        \frac{\Omega}{2}
          \left[
            A(\beta)B^2(\beta)-C(\beta)
          \right]\right.                                          
    +\left.
      \frac{1+2\Omega}{3}
        \left[
          A^3(\beta)-
            \left(\sqrt{1-
              \left(
                \frac{\chi_0+\chi_{\mathrm{b}}}%
                     {\chi_{\mathrm{b}}}\sin\beta\right)^2}
              \right)^3
        \right]
      \right\}.                                               \label{TF_15}
\end{equation}

\begin{figure}
\includegraphics[width=1.0\linewidth]{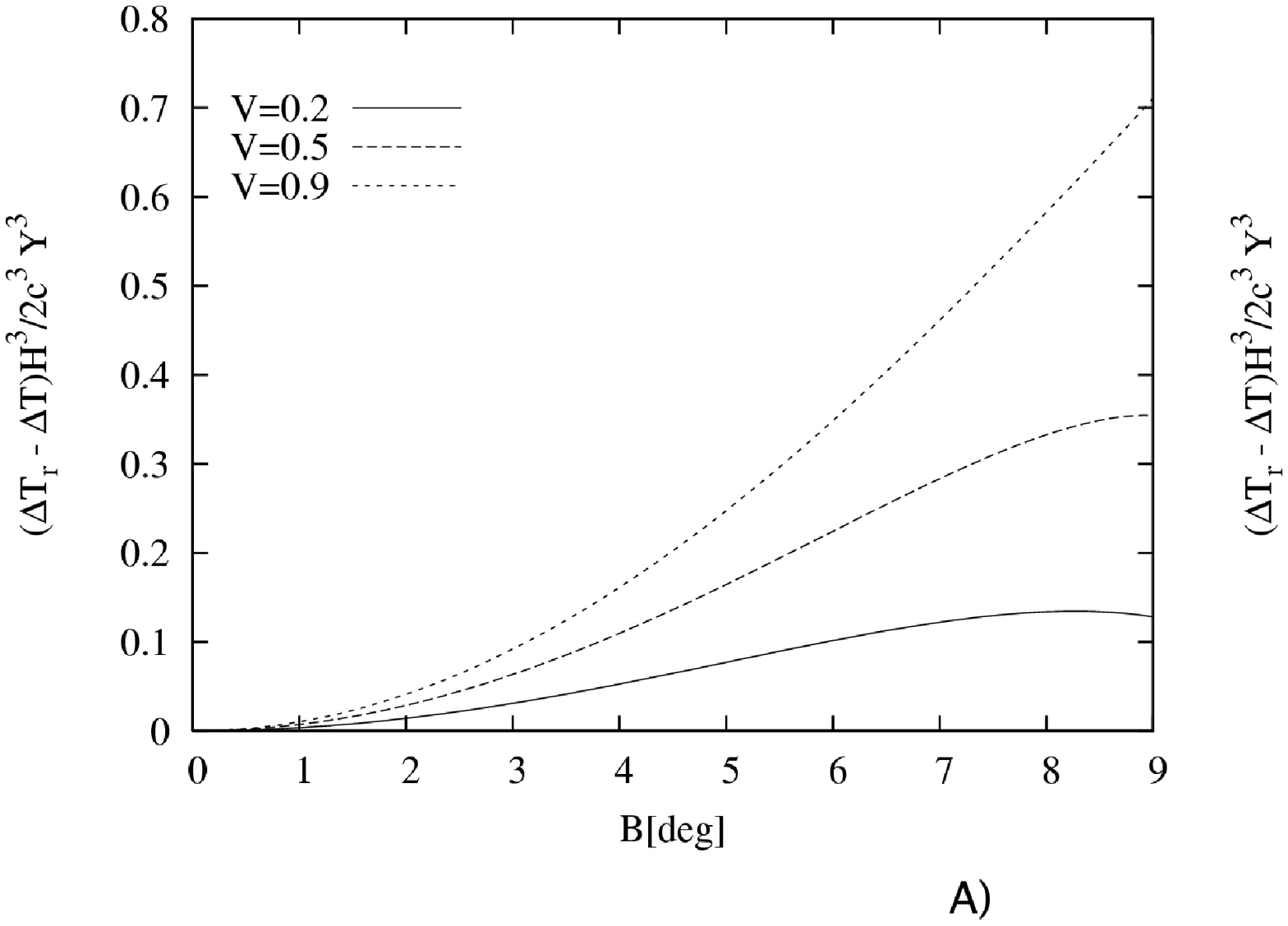}
 \caption{Plot A: Relevance of refraction is plotted as a function of angle $\beta$ for three representative values of velocity $V=0.2$, $0.5$ and $0.9$.
Plot B: Relevance of refraction is plotted as a function of velocity $V$ of MH for three representative values of angle $\beta=4^{\circ},6^{\circ}$ and $9.5^{\circ}$.
Both plots are drawn for $\Omega = 1$, $\chi_0 = 10$, $\chi_b = 2$ (see Fig\,\ref{obr3}).\label{obr4}}
\end{figure}

The relevance of the refraction effect is illustrated by Fig.\,\ref{obr4}. By analysing the relation (\ref{TF_15}), we can show that
for any value of $\beta$ the influence of the refraction on the temperature fluctuations $\Delta T_r - \Delta T$ monotonically grows with $V_r$ growing.

\section{Concluding remarks}\label{concl}
Studying the fluctuations of CMBR in an accelerating universe, we have shown, how the influence of the refraction effect 
grows with the velocity of the MH. Note that in the standard Friedman models with $\Lambda = 0$, the velocity of the MH falls 
in the expanding universe and the refraction effects are suppressed. However, in the accelerated universe, the velocity grows
after MH crosses the static radius of the SdS spacetime, and the refraction effect becomes significant. Such effect could serve as another
test of the presence of the cosmological constant; it could have strong
observational consequences in future, when the velocity of the MH becomes to be relativistic.
We conclude that there are two basic phenomena related to the importance of the refraction effect in the ESdS model 
explaining the temperature fluctuations of CMBR.

\begin{enumerate}
\item The total reflection phenomenon implies that some part of the vakuola
  region will not be visible to the external observer.  This part will be
  enlarged with expansion velocity $V_r$ growing.
\item The refraction effect on the temperature fluctuations (in the case of
  spatially flat universe) will fall, if the boundary of the MH
  $r_{\mathrm{b}}$ approaches the static radius $r_{\mathrm{s}}$ of the
  Schwarzschild-de~Sitter region, and it starts to grow after crossing the
  static radius.The effect becomes to be extremely strong when $r_{\mathrm{b}}$
  approaches the cosmological horizon $r_{\mathrm{c}}$ and $V_r\rightarrow 1$.
\end{enumerate}

We can expect that in the accelerated universe the influence of the relict
vacuum energy on the fluctuations of CMBR due to the Rees--Sciama effect could
be very important, especially the refraction effect has the tendency to rise
up the fluctuations. At present,we make our model more precise, and we
estimate conditions under which currently observable effects could be expected.

\begin{theacknowledgments}
This work was supported by the Czech grant MSM 4781305903.
\end{theacknowledgments}




\begin{thebibliography}{1}

\bibitem[Spergel et al., 2003]{Spe-etal:2003::astro-ph/0302209}
D.~N. Spergel, L. Verde, H.~V. Peiris, E. Komatsu, M.~R. Nolta, C.~L. Bennett, M.
Halpern, G. Hinshaw, N. Jarosik, A. Kogut, M. Limon, S.~S. Meyer, L. Page, G.~S. Tucker,
J.~L. Weiland, E. Wollack, and E.~L. Wright, First Year Wilkinson Microwave
Anisotropy Probe(WMAP) Observations: Determination of Cosmological Parameters. \emph{Astrophys. J. Supp. Series}, 2003, 148, Issue 1,
pp. 175--194   

\bibitem[Sachs and Wolfe, 1967]{Sac-Wol:1967:ASTRJ2:}
R.~K. Sachs and A.~M. Wolfe, \emph{Astrophys. J.}, 1967, 147:73

\bibitem[B\"{o}rner, 1993]{Bor:1993:EarlyUniv:}
G. B\"{o}rner, \emph{The Early Universe.} Springer-Verlag, Berlin-Heidelberg-New York.
1993

\bibitem[Rees and Sciama, 1968]{Ree-Sci:1968:NATURE:}
M.~J. Rees and D.~W. Sciama, \emph{Nature}, 1968, 217:511

\bibitem[M\'{e}sz\'{a}ros and Moln\'{a}r, 1996]{Mes-Mol:1996:ASTRJ2:}
A. M\'{e}sz\'{a}ros and Z. Moln\'{a}r, On the alternative origin of the dipole
anizotropy of microwave background due to the Rees--Sciama effect. \emph{Astrophys.
J.}, 1996, 470:49

\bibitem[Linde, 1990]{Lin:1990:InfCos:}
A.~D. Linde, \emph{Particle Physics and Inflationary Cosmology.} Gordon and Breach, New
York, 1990.

\bibitem[Stuchl\'{i}k, 1983]{Stu:1983:BULAI:}
Z. Stuchl\'{i}k, The motion of test particles in black-hole backgrounds with non-zero
cosmological constant. \emph{Bul. Astronom. Inst. Czechoslovakia}, 1983, 34(3):129--149.

\bibitem[Stuchl\'{i}k, 1984]{Stu:1984:BULAI:}
Z. Stuchl\'{i}k, An Einstein--Strauss--de Sitter model of the universe. \emph{Bul.
Astronom. Inst. Czechoslovakia}, 1984, 35(4):205--215.

\bibitem[Stuchl\'{i}k and Hled\'{i}k, 1999]{Stu-Hle:1999:PHYSR4:}
Z. Stuchl\'{i}k and S. Hled\'{i}k, Some properties of the Schwarzchild--de Sitter and
Schwarzchild--anti-de Sitter spacetimes. \emph{Phys. Rev. D}, 1999, 60(4):044006(15
pages).

\bibitem[Dyer and Roeder, 1973]{Dye:1973:ASTRJ2:}
C.~C. Dyer and R.~C. Roeder, \emph{Astrophys. J.}, 1973, 180:L31

\bibitem[Stuchlik and Schee, 2004]{Stu-Sch:2004:}
Z. Stuchl\'{i}k and J. Schee, \emph{Proceedings of RAGtime 4/5:Workshops on black holes and neutron stars, 14--16/13--15 October
2002/2003, Opava, Czech Republic}, 2004, p.p. 187--203

\end{thebibliography}
\end{document}